\documentclass[12pt]{article}
\usepackage{epsf}
\usepackage{graphicx,amsmath,amssymb,euscript,psfrag,epsf,epsfig}
\usepackage{cite}

\usepackage{color}
\def\beq{\begin{eqnarray}}
\def\eeq{\end{eqnarray}}

\def\Slash#1{#1 \hskip-0.6em /}

\begin{document}

\begin{titlepage}
\begin{flushright}
TUM-HEP-747/10\\
 SI-HEP-2010-03\\ 
\end{flushright}
\vskip1.cm

\begin{center}
\boldmath
{\Large\bf  Goldstone Bosons in Effective Theories \\[0.3em]
 with Spontaneously Broken Flavour Symmetry}
\unboldmath 
\vskip 2cm
{\sc M.~E.~Albrecht$^a$}\hspace*{0.1cm}, 
{\sc Th.~Feldmann$^a$}\hspace*{0.1cm}, 
{\sc Th.~Mannel$^b$}
\vskip .5cm

\vspace{0.7cm}
${}^a$ {\sf Physik Department T31, Technische Universit\"at M\"unchen,\\ 
 D-85747 Garching, Germany.}\\
\vspace{1\baselineskip}

${}^b$ {\sf Theoretische Physik 1, Fachbereich Physik,
  Universit\"at Siegen,\\ D-57068 Siegen, Germany.}\\
\vspace{3\baselineskip}

\vspace*{0.5cm}

\end{center}

\begin{abstract}
\noindent \footnotesize
The Flavour Symmetry of the Standard Model (SM) gauge sector 
is broken by the fermion Yukawa couplings. Promoting the
Yukawa matrices to scalar spurion fields, one can break 
the flavour symmetry spontaneously by giving appropriate
vacuum expectation values (VEVs) to the spurion fields, and
one encounters Goldstone modes for every broken
flavour symmetry generator. 
In this paper, we point out various aspects related to 
the possible dynamical interpretation of the Goldstone
bosons:
(i) In an effective-theory framework with \emph{local} flavour
symmetry, the Goldstone fields represent the longitudinal modes
for massive gauge bosons. The spectrum of the latter follows
the sequence of flavour-symmetry breaking related to the hierarchies
in Yukawa couplings and flavour mixing angles.
(ii) Gauge anomalies can be consistently treated
by adding higher-dimensional operators. 
(iii) Leaving the $U(1)$ factors of the flavour
symmetry group as global symmetries, the respective Goldstone
modes behave as axions which can be used to resolve the strong
CP problem by a modified Peccei-Quinn mechanism.
(iv) The dynamical picture of flavour symmetry breaking
implies new sources of flavour-changing neutral currents,
which arise from integrating out heavy scalar spurion fields and
heavy gauge bosons. The coefficients of the effective operators
follow the minimal-flavour violation principle. 
\end{abstract}

\end{titlepage}

\section{Introduction}

Flavour transitions are a very sensitive probe
for physics beyond the Standard Model (SM). Somewhat surprisingly,
high-precision tests of the Cabibbo-Kobayashi-Maskawa
matrix, in particular from $B$-meson and kaon experiments, 
so far did not find significant deviations from theoretical predictions
within the SM, although certain ``puzzles'' on the level of 2-3~$\sigma$
effects are frequently discussed and give a strong motivation for the
continuation of precision measurements in the flavour sector at future
facilities (for recent overviews, see for instance \cite{Buchalla:2008jp,Isidori:2008qp,Browder:2008em,Antonelli:2009ws}). 
As a consequence, the flavour sector of new-physics models,
that are to address the issues related to electroweak symmetry breaking and typically
contain new interactions at the TeV scale, must be very much constrained. 

A simple solution is to assume that the flavour transitions are induced
by the SM sources (i.e.\ the Yukawa matrices) only, while the new physics
at high scales is basically flavour-blind (see e.g.\ \cite{Ciuchini:1998xy,Buras:2000dm}). 
At low scales, a model-independent formulation can be achieved in terms of an
effective field theory, incorporating the concept of Minimal Flavour
Violation (MFV). Here the flavour symmetry of the SM gauge sector is considered
to be spontaneously broken by vacuum expectation values (VEVs) of spurion
fields, representing the Yukawa matrices of the SM  \cite{D'Ambrosio:2002ex}. 
In the following, we will focus on the quark sector. 
 In the lepton sector, a similar construction can be made after extending
 the minimal SM to account for neutrino masses and mixing angles  
  \cite{Cirigliano:2005ck}.

Taking this framework at face value, one has to promote the
entries of the Yukawa matrices to complex scalar fields which
become dynamical at high scales $\Lambda$ 
and are embedded into an effective theory with a global flavour symmetry. 
It then appears rather natural to
assume that the hierarchy of fermion masses and mixing angles in the SM
is related to a hierarchy of scales at which the different scalar
field components in the Yukawa matrices acquire a VEV. For instance,
the ${\cal O}(1)$-value of the top-quark Yukawa coupling suggests that
the VEV of the corresponding scalar field is generated already at
the high scale $\Lambda$, while the remaining quark Yukawa
couplings are generated at lower scales $\Lambda_i \ll \Lambda$.
Below $\Lambda$, the effective theory
is conveniently formulated by representing the flavour symmetry in
a non-linear way, including Goldstone fields -- for that part of the
flavour symmetry that is broken by the top-quark Yukawa coupling --
and residual spurion fields for the unbroken symmetry \cite{Feldmann:2008ja}
(see also \cite{Kagan:2009bn}).
An analogous construction can be successively applied to the next-largest
entries in the Yukawa matrices, and the resulting sequence of partly broken
flavour symmetries -- which reflects the hierarchies of quark masses and
mixing angles in the SM -- can be identified \cite{Feldmann:2009jung}.
A similar framework can be defined for the lepton sector \cite{Feldmann:2008av},
for instance in a minimally extended SM where the neutrino masses are determined by
a dimension-5 operator.

There have been many attempts to
explain the hierarchies in the quark and lepton Yukawa sector
by enforcing certain
flavour symmetries,  often in the framework of 
(more fundamental) extensions of the SM. The
simplest example has been proposed by Froggatt and Nielsen \cite{Froggatt:1978nt},
where different $U(1)$ charges are assigned to the different fermion
generations, and the hierarchies of masses and mixing angles are
generated by higher-dimensional operators involving different powers
of a scalar field charged under the new $U(1)$. Refined extensions
of this idea have been discussed by many authors,
in particular in the context of supersymmetric and/or grand unified scenarios 
\cite{Leurer:1992wg,Frampton:1994rk,Barbieri:1998em,Berezhiani:2000cg,King:2003rf,Chen:2003zv,Antusch:2007re,Stech:2008wd}.

In this paper,  we pursue the effective-theory construction suggested by the MFV framework,
where the scalar sector consists of the SM Higgs field together with the Yukawa spurions.
We then focus on the 
dynamical interpretation of the
Goldstone modes related to the broken flavour symmetries.\footnote{
An alternative picture, where
the concept of MFV is applied to \emph{discrete} sub-groups of the
flavour symmetry, has been discussed in \cite{Zwicky:2009vt}. 
In such a context, Goldstone modes would not appear when the discrete 
flavour symmetry is spontaneously broken.}
More specifically, we consider two alternative cases:
\emph{local} flavour symmetries, where
as usual the Goldstone modes become the longitudinal modes of the
massive gauge bosons associated with the broken symmetry generators,
and ``invisible-axion'' scenarios, where the Goldstone modes are
(almost) massless fields with ultra-weak couplings to the SM fermions.

The main purpose of the subsequent analysis is to 
understand whether such a scenario can be consistently defined,
and to identify the potential new sources for flavour-violating 
processes at low energies. Among others, this requires to address
the following issues:
(i) 
The hierarchies of quark Yukawa couplings and mixings
imply a sequence of flavour-symmetry breaking \cite{Feldmann:2009jung}
which will be reflected in the mass spectrum of the associated
massive gauge bosons. Integrating out the heavy gauge bosons leads to
effective four-quark operators at tree-level, which potentially
induce new flavour-changing currents. The strongest effect is expected
from the lightest of the heavy gauge bosons which correspond to the
latest stages in the sequence of flavour-symmetry breaking. 
To illustrate the type of flavour-violating effects, it will therefore
be sufficient to perform the analysis for the smallest non-trivial 
flavour symmetry sub-groups identified in \cite{Feldmann:2009jung}.
(ii)
Restricting ourselves to the SM field content for fermion matter
fields, we naturally encounter gauge anomalies \cite{Wess:1971yu},
when promoting the global flavour symmetries to local ones.
As is well-known, local gauge invariance can be formally 
restored by adding appropriate higher-dimensional operators,
following, for instance, the general framework worked out 
by Preskill in \cite{Preskill:1990fr}. The higher-dimensional operators
can be considered as the result of integrating out new heavy fermion modes
which receive masses of the order of the cut-off scale in the corresponding effective
theory with (partly) broken local flavour symmetry.
(iii)
In addition to the Goldstone modes, we have to take into account 
the fluctuations around the VEVs for the spurion fields,
representing physical scalar Higgs fields of the broken flavour symmetry.
Again, these fields are expected to develop a hierarchical mass spectrum
following the sequence of scales identified in  \cite{Feldmann:2009jung},
and corresponding flavour-changing 4-quark operators can be induced
at tree level.
A technical complication arises due to the mixing of Goldstone
modes in the non-Abelian theory and the spurion fields related to the 
mixing angles between different families. In order to identify the physical
couplings between the flavoured Higgs fields and the quark currents, we thus
have to find the correct prescription for the unitary-gauge condition.
(iv)
A special role is played by the Abelian $U(1)$ factors in
the SM flavour group (see below), which can be considered
as Peccei-Quinn (PQ) symmetries \cite{Peccei:1977ur}. 
As a benefit of our framework, we can leave the $U(1)$ factors
as global symmetries, such that the corresponding
Goldstone mode(s) become dynamical axion fields, realizing  a 
variant  of the standard PQ solution to the strong CP problem.


\section{Local Flavour Symmetries}

The global flavour symmetry group in the SM quark sector (consisting of
unitary transformations that leave the gauge sector invariant, but are
broken by the Yukawa couplings) is given by\footnote{Notice that one
combination of $U(1)$ phases corresponds to baryon number that is not
broken by the Yukawa matrices and thus should not be considered part of
the flavour symmetry. Furthermore, the phase related to
weak hypercharge transformations should be related to unitary transformations
of the Higgs field and therefore does not appear in $G_F$, see
\cite{Berger:2008zq,Feldmann:2009jung}.} 
\begin{align}
 G_F & = SU(3)_{Q_L} \times SU(3)_{U_R} \times SU(3)_{D_R} \times U(1)_{U_R} \times U(1)_{D_R}\,,
\label{GF:original}
\end{align}
where the subscripts refer to the left- and right-handed gauge multiplets in the SM.
In the course of spontaneous flavour symmetry breaking each scalar spurion contained in the Yukawa matrices $Y_U$ and $Y_D$ acquires a VEV. 
Without loss of generality we may choose to work in the basis, where the VEV of the up-type Yukawa matrix is diagonal.
\begin{align}
 \left\langle Y_U \right\rangle&=\mathrm{diag}(y_u \, e^{-i\pi_u},y_c,y_t) \,,
\qquad 
 \left\langle Y_D \right\rangle= V_{\rm CKM} \, \mathrm{diag}(y_d \, e^{-i\pi_d}, y_s,y_b) \,.
\label{Yvev}
\end{align}
The CKM matrix can be written as  
\begin{align}
 V_{\rm CKM} &=  \exp\left[ 2 i \theta_{23} \, T^7\right] \, 
  \exp\left[2 i \theta_{13} \, T^5\right] \,
  \exp\left[i \delta \, T^3\right]  \exp\left[2 i \theta_{12} \, T^2\right] 
 \exp\left[ -i \delta \, T^3 \right] \,,
\label{Vckm}
\end{align}
which corresponds to the standard parametrization up to a re-phasing,
 such that the CP-phase $\delta$ appears in the mixing between
the first and second generation.

\subsection{Massive Gauge Bosons}

In the following, we will consider the specific scenario where the three $SU(3)$
factors will be promoted to \emph{local} symmetries, 
while keeping the two $U(1)$ factors in $G_F$ as  global symmetries.
The Goldstone modes related to the latter have anomalous couplings with the SM gauge fields,
and we expect that one linear combination of the associated Goldstone bosons
will contribute to the effective $\theta$\/--parameter in QCD. It can
thus be identified as an axion providing a potential solution to the strong CP-problem,
with a finite mass generated by anomalous couplings to QCD instantons.
On the other hand, upon spontaneous symmetry breaking, the 24 gauge bosons of $SU(3)^3$
will become massive, and (in unitary gauge)  the Goldstone modes of the broken flavour symmetry
shall be identified with the longitudinal modes of massive gauge-boson. 
These masses are generated, as usual, from
the gauge-kinetic term for the spurion fields,
\begin{align}
 \Lambda^2 \, {\rm tr} \left[ (D_\mu Y_U^\dagger)(D^\mu Y_U) \right]
+\Lambda^2 \, {\rm tr} \left[ (D_\mu Y_D^\dagger)(D^\mu Y_D) \right] \,,
\label{kin}
\end{align}
which we have normalized to a UV scale $\Lambda$.
This yields a $24\times 24$ mass matrix for the gauge fields $A^{\mu a}_{Q_L}$,  $A^{\mu a}_{U_R}$, 
$A^{\mu a}_{D_R}$,  when $Y_{U,D}$ are replaced by their VEVs. 
To diagonalize this matrix, it is convenient to take advantage of the sequence of symmetry breakings.
For concreteness, let us assume 
a particular realization for a sequence of flavour symmetry breaking \cite{Feldmann:2009jung},
$$
   y_t > y_b > y_c > y_b \theta_{23} > y_b \theta_{13} > y_s > y_s \theta_{12} > y_{u,d} \,,
$$
as a reference case.
The first gauge bosons to achieve a mass from spontaneous symmetry breaking are thus
the ones related to the 9 broken generators associated to the breaking of 
$SU(3)^3 \to SU(3) \times SU(2)^2 \times U(1)$ through a non-vanishing value for
the top Yukawa coupling $y_t$. In this case, the diagonalization of the mass matrix is
trivial and yields:
\begin{align}
 W_{Q_L}^{(13,31)} \,, \ W_{Q_L}^{(23,32)} & : \quad M^2 = \frac12 \, g_{Q_L}^2 y_t^2\Lambda^2 \,, \cr 
 W_{U_R}^{(13,31)} \,, \ W_{U_R}^{(23,32)} & : \quad M^2 =  \frac{\cot^2 \phi_1}{2} \, g_{Q_L}^2  y_t^2 \Lambda^2 \,, \cr
 Z_1 = -\sin\phi_1 \, A_{Q_L}^8 + \cos\phi_1 \, A_{U_R}^8 & : \quad M^2 =  \frac{2 }{3 \sin^2\phi_1} \, g_{Q_L}^2  y_t^2 \, \Lambda^2 \,, \cr 
 A_1 =  \cos \phi_1 \, A_{Q_L}^8 + \sin\phi_1 \, A_{U_R}^8 & : \quad M^2 =  0 \,,
\end{align}
with $\tan\phi_1 = g_{Q_L}/g_{U_R}$, and all other eigenvalues being zero.

The mixing of the heaviest gauge bosons (with $M \propto g y_t \Lambda$) and the remaining gauge bosons only
occurs from the next symmetry-breaking steps and is thus suppressed by at least $y_b/y_t \ll 1$. In order
to discuss the effect of the next breaking, $SU(3)\times SU(2)^2 \times U(1) \to SU(2)^3 \times U(1)$,
--- to first approximation --- we can therefore simply neglect the presence of the heaviest gauge bosons and
diagonalize the mass matrix for the 5 next heaviest gauge bosons, yielding
\begin{align}
 W_{D_R}^{(13,31)} \,, \ W_{D_R}^{(23,32)} & : \quad M^2 =  \frac{\cot^2\phi_2}{2} \, (g_{Q_L} \cos\phi_1)^2 y_b^2 \Lambda^2 \,, \cr 
 Z_2 = -\sin\phi_2 \, A_1 + \cos\phi_2 \, A_{D_R}^8 & : \quad M^2 = \frac{2}{3\sin^2\phi_2} \, (g_{Q_L} \cos\phi_1)^2 y_b^2 \Lambda^2 \,, \cr 
 A_2 = \cos\phi_2 \, A_1 + \sin\phi_2 \, A_{D_R}^8 & : \quad M^2 = 0 \,,
 \end{align}
with $\tan\phi_2 = g_{Q_L} \cos\phi_1/g_{D_R}$.

This scheme can be continued until the complete flavour symmetry is broken, and the
hierarchies of Yukawa couplings and mixing angles translates into a hierarchy of gauge boson masses:
\begin{align}
\begin{matrix}
 && y_t & \gg y_b & \gg y_c  & \gg y_b \theta_{23} &  \gg y_b \theta_{13} & \gg y_s & \gg y_s \theta_{12} & \gg y_{u,d} \,,
 \\[0.8em]
 M \ : && g\Lambda & \gg g\Lambda'   & \gg g\Lambda''  & \gg g\Lambda'''  & \gg g\Lambda^{(iv)}
 & \gg g\Lambda^{(v)}  & \gg g\Lambda^{(vi)} & \gg g\Lambda^{(vii)}  \,,
 \end{matrix} 
\end{align}
with $\Lambda'/\Lambda = y_b/y_t$ etc.~\cite{Feldmann:2009jung}.


In addition to the spontaneous symmetry breaking from the spurion VEVs, the
presence of gauge anomalies has to be taken care of by adding appropriate 
higher-dimensional operators to formally restore invariance under local 
symmetry transformations. This can be achieved by applying the general formalism
developed in \cite{Preskill:1990fr}: 
At every step in the sequence of effective
theories, we can imagine to add appropriate fermion representations under the
flavour symmetry group to cancel the gauge anomalies of $SU(3)^3$ or its sub-groups.
In the simplest case, we could just introduce chiral partners for the SM fermions,
\begin{align}
Q_L  &\leftrightarrow (\tilde U_R,\tilde D_R) \,, \qquad
U_R  \leftrightarrow \tilde U_L \,, \qquad
D_R  \leftrightarrow \tilde D_L \,,
\end{align}
in the same flavour representations, but with vanishing $SU(2)_L$ quantum numbers, and hypercharges
taken as $y_{\tilde U_R}=y_{\tilde U_L}=y_{U_R}$, $y_{\tilde D_R}=y_{\tilde D_L}=y_{D_R}$ 
which also satisfy $y_{\tilde U_R}+y_{\tilde D_R}=2 y_{Q_L}$. Concerning the colour group, the new
quarks can be taken as 3-fold degenerate colour singlets, such that the mixed anomalies between
$U(1)_{U_R+D_R}$ and the QCD generators remains as in the SM.\footnote{Notice that, necessarily, the new fermions
have to be singlets under the chiral $SU(2)_L$ group, because they have to receive masses \emph{before} 
electroweak symmetry breaking. As a consequence, also the mixed anomaly between baryon number and $SU(2)_L$
remains as in the SM. In order to gauge baryon number, one could, for instance, add a fourth generation
with different gauge quantum numbers along the lines of \cite{FileviezPerez:2010gw}.}
These fermions
get their masses directly from the Yukawa spurion VEVs via couplings like $\bar {\tilde U}_R \, Y_U \, \tilde U_L$ 
and $\bar {\tilde D}_R \, Y_D \, \tilde D_L$. For instance, when $Y_U$ develops a VEV related to the top-Yukawa
coupling, the corresponding heavy fermion $\tilde T$ gets a mass of the order of the UV scale $\Lambda$
in the effective theory. Integrating out the auxiliary fermion fields, on the one hand,
leaves the remaining fermion representations anomalous under the flavour symmetry group, but
at the same time generates the higher-dimensional operators compensating for the change in
the path-integral under local chiral rotations.
Insertion of these higher-dimensional operators
into loop diagrams in general induces additional contributions to the 
gauge-boson masses proportional to the respective anomaly coefficients.

\subsection{Simplified Set-Up}

In the following, we will assume that the gauge-boson masses
are dominated by spontaneous symmetry breaking. The most important effects
for flavour phenomenology will thus come from the lightest of these gauge bosons,
which then correspond to the smallest non-trivial sub-group of 
$SU(3)^3$ which, in the above example, is given by \cite{Feldmann:2009jung}
\begin{align}
 G_F^{(4)} = SU(2)_{D_R} \times U(1)_X 
\label{eq:G4}
\end{align}
Here, in contrast to (\ref{GF:original}),
$D_R =(d_R,s_R)$ is restricted to a right-handed flavour doublet of
down-type quarks.
In terms of the symmetry generators of the original flavour symmetry in (\ref{GF:original}),
the local $U(1)_X$ charge is given by\footnote{In contrast to \cite{Feldmann:2009jung},
we are not allowed to consider linear combinations with the remaining \emph{global}
symmetry generators (including baryon number).}
\begin{align}
  Q_X &= \frac{1}{\sqrt3} \left( T^8_{Q_L}+T^8_{U_R}+T^8_{D_R} \right)
      + \left( T^3_{Q_L} + T^3_{U_R} \right) \,.
\label{eq:Xdef}
\end{align}
This linear combination leaves the VEVs of the Yukawa matrices 
$$
 \langle Y_U \rangle = \left(
  \begin{matrix}  0 & 0 & 0 \\ 0 & \bullet & 0 \\ 0 & 0 & \bullet \end{matrix}
  \right)\,, \qquad
\langle Y_D\rangle = \left(
  \begin{matrix} 0 & 0 & 0 \\ 0 & 0 & \bullet \\ 0 & 0 & \bullet \end{matrix}
  \right)
 = \exp\left[-i\left(
  \begin{matrix} 0 & 0 & 0 \\ 0 & 0 & \bullet \\ 0 & \bullet & 0 \end{matrix}
  \right)\right] \cdot \left(
  \begin{matrix} 0 & 0 & 0 \\ 0 & 0 & 0 \\ 0 & 0 & \bullet \end{matrix}
  \right) \,,
$$
invariant, where the bullet denotes a non-zero entry.\footnote{\label{footX}
Notice the different treatment of up- and down-type quarks
in the definition of the charge operator $Q_X$, which is correlated with the particular
choice for the parametrization of the Yukawa matrices, where $\langle Y_U\rangle $
is diagonal, while  $\langle Y_D \rangle$ contains the CKM rotations.}
The above Yukawa structures 
correspond to the situation where 3 eigenvalues  ($y_t,y_b,y_c$)
and the mixing angle between the $2^{nd}$ and $3^{rd}$ generation are non-zero. 

The local $U(1)_X$ symmetry in (\ref{eq:Xdef}) 
has an anomaly that will be treated by the formalism of \cite{Preskill:1990fr}, see Appendix~\ref{app:U1X}.
The local $SU(2)_{D_R}$ symmetry has no anomalies, and the corresponding
gauge theory can be written down in a straight-forward manner. 
The symmetry $G_F^{(4)}$ will be further broken down to
$$
G_F^{(4)} \longrightarrow G_F^{(5)} = SU(2)_{D_R} \longrightarrow \mbox{nothing} \,,
$$
and the corresponding spurions and massive gauge fields have to
be integrated out, inducing new flavour structures for higher-dimensional
quark operators.
After this,
we are left with only global (anomalous) Abelian symmetries. 
Their breaking involves physical Goldstone modes which, in our case,
will provide a solution to the strong CP problem similar to the usual
axion scenario.


\section{Spurion Fields and Unitary Gauge}



\label{ug}

From the $4n^2$ real fluctuations of the two complex Yukawa matrices
(for $n=3$ quark generations) we shall identify $3(n^2-1)$ as Goldstone bosons
$\phi_{Q_L,U_R,D_R}^a(x)$
of $SU(n)^3$, together with $(n^2+1)$ fluctuations $\eta_i(x)$ of the physical masses
and mixing parameters, and 2 explicit phases for the global $U(1)_{U_R}\times U(1)_{D_R}$
symmetry. 

The first non-trivial task is to identify the unitary (i.e.\ physical) 
gauge, where the massive $SU(3)^3$ gauge bosons do not mix with the 
fluctuations $\eta_i(x)$. For this purpose we introduce the 
parametrization
\begin{align}
 Y_U(x) &= \Sigma_{Q_L}(x) \cdot \Xi_{U_L}(x) \, D_U(x) \, \Xi_{U_R}^\dagger(x) \cdot 
    \Sigma_{U_R}(x) \,,
\\
 Y_D(x) &= \Sigma_{Q_L}(x) \cdot V_{\rm CKM} \cdot
    \Xi_{D_L}(x) \, D_D(x) \, \Xi_{D_R}^\dagger(x) \cdot 
    \Sigma_{D_R}(x) \,.
\label{ansatz}
\end{align}
Here, 
\begin{align}
 D_U(x) &= \mathrm{diag}\left(
  y_u(x) \,  e^{-i \pi_u(x)}  ,\
  y_c(x) ,\
  y_t(x) \right) \,,
\cr 
 D_D(x) &= \mathrm{diag}\left(
  y_d(x) \, e^{-i \pi_d(x)}  ,\
  y_s(x) ,\
  y_b(x) \right)
\end{align}
are diagonal matrices, 
with $y_i(x) = y_i + \eta_i(x)/\sqrt2$, 
containing the fluctuations around the VEVs for the quark Yukawa couplings, and
\begin{align}
 \Sigma_{X}(x) &= \exp\left[ i \, \phi_{X}^a(x) \, T^a \right] \,, \qquad (X=Q_L,U_R,D_R)
\end{align}
represent the Goldstone bosons of $SU(3)^3$, while 
\begin{align}
 \Xi_X(x) &= \exp\left[ i \, \xi_{X}^a(x) \, T^a \right] \,, \qquad (X=U_L,D_L,U_R,D_R;\
 a \neq 3,8)
\end{align}
parameterize the scalar fluctuations around the VEVs for CKM angles and phases.
Notice that our ansatz contains more parameters than scalar degrees of freedom.
The apparent ambiguity is resolved by \emph{requiring} that in
\begin{align}
\mbox{unitary gauge, i.e.} \quad \Sigma_X(x) & \to 1 \,,
\end{align}
our ansatz does not generate mixing terms with the $SU(3)^3$ gauge bosons, when
inserted into the gauge-invariant kinetic term (\ref{kin}),
where the covariant derivatives read
\begin{eqnarray}
 D_{\mu} Y_U(x) &=&
\partial_{\mu} Y_U(x)
-i g_{Q_L} \, A_{Q_L,\mu}^a(x) T^a \, Y_U(x)
+i g_{U_R} \, A_{U_R,\mu}^a(x) \, Y_U(x) \, T^a 
\,, \cr 
D_{\mu} Y_D(x) &=&
\partial_{\mu} Y_D(x)
-i g_{Q_L} \, A_{Q_L,\mu}^a(x) T^a \, Y_D(x)
+i g_{D_R} \, A_{D_R,\mu}^a(x) \, Y_D(x) \, T^a 
\,,
\end{eqnarray}
and $A_{X,\mu}^a(x)$ are the gauge fields for $SU(3)^3$.
Here $\Lambda$ is the high-energy scale related to the ultra-violet (UV)
cut-off of the effective theory.
Inserting the parametrization (\ref{ansatz}) and focusing on the mixing
terms between the scalar fluctuations $\xi_i(x)$ and 
the gauge fields, we obtain the set of conditions
\begin{align}
A^\mu_{U_R} &: \quad
 \mathrm{tr}\left[T^a \left( 
\Xi_{U_R} \langle D_U^\dagger\rangle (i\partial_\mu \Xi_{U_L}^\dagger) \Xi_{U_L} 
   \langle D_U \rangle  \Xi_{U_R}^\dagger 
+ (i\partial_\mu \Xi_{U_R}) \langle|D_U|^2\rangle \Xi_{U_R}^\dagger + \mbox{h.c.}
\right) \right] =0 \,,
\label{cond1}
\\[0.2em] 
A^\mu_{D_R} &: \quad
 \mathrm{tr}\left[T^a \left( 
\Xi_{D_R} \langle D_D^\dagger \rangle (i\partial_\mu \Xi_{D_L}^\dagger) \Xi_{D_L} 
   \langle D_D \rangle  \Xi_{D_R}^\dagger 
+ (i\partial_\mu \Xi_{D_R}) \langle |D_D|^2 \rangle \Xi_{D_R}^\dagger + \mbox{h.c.}
\right) \right] =0 \,,
\label{cond2}
\end{align}
and
\begin{align}
& 0=
\mathrm{tr}\left[T^a \left( 
\Xi_{U_L} \langle |D_U|^2 \rangle (i\partial_\mu \Xi_{U_L}^\dagger) 
+ \Xi_{U_L} \langle D_U\rangle \Xi_{U_R}^\dagger (i\partial_\mu \Xi_{U_R}) 
 \langle D_U^\dagger \rangle \Xi_{U_L}^\dagger + \mbox{h.c.}
\right) \right.
\cr 
& \left.
{} + V_{\rm CKM}^\dagger T^a V_{\rm CKM}
\left( 
\Xi_{D_L} \langle |D_D|^2 \rangle (i\partial_\mu \Xi_{D_L}^\dagger) 
+ \Xi_{D_L} \langle D_D\rangle \Xi_{D_R}^\dagger (i\partial_\mu \Xi_{D_R}) 
 \langle D_D^\dagger \rangle \Xi_{D_L}^\dagger + \mbox{h.c.}
\right)\right] 
\cr & 
\label{cond3}
\end{align}
for $A^\mu_{Q_L}$.
(The fluctuations around the quark Yukawa couplings in $D_U$ and $D_D$
do not mix with the $SU(3)^3$ gauge bosons and can be dropped for this part
of the analysis.)

\subsection{Approximation for the 3-Family Case}

In the 2-family case, the above equations can be solved explicitly, 
see Appendix~\ref{app:unitary2}.
In the 3-family case, the situation is somewhat more complicated because
the generators for 3 CKM rotations do not commute anymore, and the 
kinetic terms for the related spurion fields 
$\eta_{12}(x)$, $\eta_{13}(x)$ and $\eta_{23}(x)$ will also mix,
with the mixing controlled by the CKM matrix and the ratios of
quark Yukawa couplings.
We can identify the leading effects by expanding to
first order in the off-diagonal CKM elements $V_{i\neq j}$,
and assuming  $y_s \ll y_b$ and $y_c \ll y_t$.
For simplicity, we will also set $y_u=y_d=0$ and neglect the CP-phase
in the CKM matrix.
With these approximations, we obtain
\begin{align}
& Y_U^{\rm u.g.}(x)  \simeq 
{\rm diag}\left[0,y_c,y_t \right]
 + 
 \left( \begin{array}{ccc}
0 & \frac{1}{y_c}   &
    - \frac{\theta_{23} y_b^2 y_t}{y_b^2y_c^2-y_s^2y_t^2}
  \\[0.3em]
0 & 0 & \frac{\theta_{13} y_b^2 y_t}{y_b^2y_c^2-y_s^2y_t^2}
 \\[0.3em]
0  & \frac{\theta_{13} \, y_b^2y_c}{y_b^2 y_c^2-y_s^2 y_t^2}   & 0
\end{array} \right) 
\frac12 \, F_{12}^2 \, \tilde\eta_{12}(x)
\cr 
& \quad 
+ \left( \begin{array}{ccc}
0 & \frac{\theta_{23} y_b^2 y_c }{y_b^2 y_c^2-y_s^2 y_t^2}   &
    \frac{1}{y_t}
  \\[0.3em]
0 & 0 & \frac{\theta_{12} y_s^2 y_t}{y_b^2 y_c^2 - y_s^2 y_t^2}
 \\[0.3em]
0  &  0  & 0
\end{array} \right) 
\frac12 \, F_{13}^2 \, \tilde \eta_{13}(x)
+
\left( \begin{array}{ccc}
0 & - \frac{\theta_{13} y_b^2 y_c}{y_b^2 y_c^2-y_s^2 y_t^2} 
  & - \frac{\theta_{12} y_s^2 y_t}{y_b^2 y_c^2-y_s^2 y_t^2}
  \\[0.3em]
0 & 0 & \frac{1}{y_t}
 \\[0.3em]
0  &  0   & 0
\end{array} \right) 
\frac12 \, F_{23}^2  \,\tilde \eta_{23}(x) 
 \,,
\cr 
\label{YUapprox}
\end{align}
and
\begin{align}
& V_{\rm CKM}^\dagger \, Y_D^{\rm u.g.}(x)  \simeq 
{\rm diag}\left[0,y_s,y_b \right]
 + 
 \left( \begin{array}{ccc}
0 & -\frac{1}{y_s}   &
    \frac{\theta_{23} y_b y_t^2}{y_b^2y_c^2-y_s^2y_t^2}
  \\[0.3em]
0 & 0 & - \frac{\theta_{13} y_b y_t^2}{y_b^2y_c^2-y_s^2y_t^2}
 \\[0.3em]
0  & -\frac{\theta_{13} \, y_s y_t^2}{y_b^2 y_c^2-y_s^2 y_t^2}   & 0
\end{array} \right) 
\frac12 \, F_{12}^2 \, \tilde\eta_{12}(x)
\cr 
& \quad 
+ \left( \begin{array}{ccc}
0 & -\frac{\theta_{23} y_s y_t^2  }{y_b^2 y_c^2-y_s^2 y_t^2}   &
    -\frac{1}{y_b}
  \\[0.3em]
0 & 0 & - \frac{\theta_{12} y_b y_c^2 }{y_b^2 y_c^2 - y_s^2 y_t^2}
 \\[0.3em]
0  &  0  & 0
\end{array} \right) 
\frac12 \, F_{13}^2 \, \tilde \eta_{13}(x)
+
\left( \begin{array}{ccc}
0 &  \frac{\theta_{13} y_s y_t^2 }{y_b^2 y_c^2-y_s^2 y_t^2} 
  &  \frac{\theta_{12} y_b y_c^2}{y_b^2 y_c^2-y_s^2 y_t^2}
  \\[0.3em]
0 & 0 & -\frac{1}{y_b}
 \\[0.3em]
0  &  0   & 0
\end{array} \right) 
\frac12 \, F_{23}^2  \,\tilde \eta_{23}(x) 
 \,.
\cr 
\label{YDapprox}
\end{align}
The normalization factors $F_{ij}$ are defined as in Appendix~\ref{app:unitary2},
\begin{align}
F_{ij}^2 & = \frac{2 (y_{U_i}^2-y_{U_j}^2)^2 (y_{D_i}^2-y_{D_j}^2)^2}{(y_{U_i}^2-y_{U_j}^2)^2
  (y_{D_i}^2+y_{D_j}^2) + (y_{D_i}^2-y_{D_j}^2)^2 (y_{U_i}^2+y_{U_j}^2)}  \,.
\end{align}
We have brought the result into a form where up- and down-quarks couplings are treated symmetrically, and
the fluctuations $\delta Y_U$ around the VEV satisfy
\begin{align}
 {\rm tr}\left[\langle Y_U^\dagger\rangle \, \delta Y_U \right]
 = {\rm tr}\left[\langle Y_U^\dagger Y_U Y_U^\dagger \rangle \, \delta Y_U \right]
 = {\rm tr}\left[\langle Y_U^\dagger Y_U Y_U^\dagger Y_U Y_U^\dagger\rangle \, \delta Y_U \right] = 0 \,,
\end{align}
and analogously for $\delta Y_D$.
Furthermore, the contribution to the invariants ${\rm tr}[(Y_U^\dagger Y_U)^n]$,
appearing in the spurion potential \cite{Feldmann:2009jung} are diagonal in
the spurion fields $\tilde \eta_{ij}(x)$. 

In addition to the sub-leading terms in the kinetic mixing terms,
further corrections would be induced by radiative
corrections involving the Yukawa couplings.
By construction, these effects will follow the MFV principle, 
in a similar way as we have discussed for the 2-family
example in Appendix~\ref{app:unitary2}. 
A precise calculation of these terms is beyond the scope of this work.

\section{Flavour Transitions from Local $SU(2)_{D_R} \times U(1)_X$}

\label{sec:preskill}

\subsection{Effective Lagrangian}

The Lagrangian for our effective theory with a local flavour symmetry
$G_F^{(4)}=SU(2)_{D_R} \times U(1)_X$ is constructed from the SM matter and
gauge fields, together with scalar spurions and gauge fields for $G_F^{(4)}$.
The various left-handed fermions
build irreducible representations of $G_F^{(4)} \times \mbox{SM}$,
\begin{align}
\psi_L:  & \quad \{ Q_L^{(1)}, Q_L^{(2)}, Q_L^{(3)}, u_R^c,\,\, c_R^c,\,\, t_R^c,\,\,
 D_R^{c},\ b_R^c \} \label{psiL}
\\
 q_X: & \quad \nonumber \{ +\frac23,\ -\frac13 ,\ -\frac13 ,\ -\frac23 ,\ +\frac13 , +\frac13 ,\ -\frac16,\ +\frac13 \}  \,,
\end{align}
where in the  last line we denoted the respective charges under $U(1)_X$. 
The SM Lagrangian takes its standard form, except for the Yukawa sector,
where we have to replace the Yukawa matrices by dynamical fields.
Starting from the general discussion in Sec.~\ref{ug}, we can drop all
Goldstone modes and spurion fields related to the breaking $G_F \to G_F^{(4)}$,
and obtain
\begin{align}
 Y_U \to Y_U(x) &= \ e^{i \pi_X(x) (T^3 +T_8/\sqrt 3)} 
\cdot \tilde Y_U^{\rm u.g.}(x) \cdot e^{- i \pi_X(x) (T^3 +T_8/\sqrt 3)}
 \,,
\cr
 Y_D \to Y_D(x) &= \ e^{i \pi_X(x) (T^3 +T_8/\sqrt 3)} \cdot 
  \tilde Y_D^{\rm u.g.}(x) \cdot  e^{-i \sum\limits_{a=1}^3 \pi_{D_R}^a(x) T^a }
\, e^{- i \pi_X(x) T_8/\sqrt 3}
\,
\end{align}
where $\tilde Y_{U,D}^{\rm u.g.}(x)$ are given by Eqs.~(\ref{YUapprox},\ref{YDapprox}) 
with $\tilde\eta_{23}(x)$ set to zero.
The Lagrangian is supplemented by a spurion kinetic and potential term
\begin{align}
 {\cal L}_{\rm spurion} & = \ \Lambda^2 \, {\rm tr} \left[(D^\mu Y_U^\dagger) (D_\mu Y_U) \right]
+ \Lambda^2 \, {\rm tr} \left[(D^\mu Y_D^\dagger) (D_\mu Y_D) \right]
- V(Y_U, Y_D)
\label{Lspurion}
\end{align}
with 
\begin{align}
  D_\mu Y_U(x) & = \ \partial_\mu Y_U(x) 
  - i g_X  X_\mu(x) \left[ T^3+T^8/\sqrt3, \, Y_U(x) \right]
\,,
\cr 
  D_\mu Y_D(x) & = \ \partial_\mu Y_D(x) 
  + i g_{D_R} \sum_{a=1}^{3} \, A_{D_{R},\mu}^{a}(x) \, Y_D(x) \, T^a_{D_R}
\cr & \quad 
  - i g_X X_\mu(x) \, (T^3+T^8/\sqrt3) \, Y_D(x)
  + i g_X X_\mu(x) \, Y_D(x) \, T^8/\sqrt3 \,,
\end{align}
and a $G_F^{(4)}$-invariant potential $V(Y_U,Y_D)$.
Inserting the above representation for $Y_U$ and $Y_D$, expanding
in the gauge and spurion fields, and using again the approximations
in terms of small Yukawa couplings and CKM angles, we identify the 
kinetic terms for the spurion fields, as well as
the mass terms for the $SU(2)_{D_R} \times U(1)_X$ gauge bosons induced by the
VEVs for $\theta_{13}$, $\theta_{12}$ and $y_s$:
\begin{align}
{\cal L}_{\rm kin} &=  
\Lambda^2 \left(\partial_\mu y_s(x)\right)^2  
+
\frac{\Lambda^2}{2} \, F_{12}^2 \, \left(\partial_\mu \tilde\eta_{12}(x)\right)^2
+
\frac{\Lambda^2}{2} \, F_{13}^2 \, \left(\partial_\mu \tilde\eta_{13}(x)\right)^2 \,,
\label{Lgen1}
\\
{\cal L}_{\rm mass} &\simeq 
\Lambda^2 
 y_b^2 \theta_{13}^2 
\left( \partial_\mu \pi_X - g_X X_\mu \right)^2
\cr 
& \quad 
+
\frac{\Lambda^2}{4} \left( y_s(x) \right)^2 
\left[ 2 \, {\cal A}_\mu^+ {\cal A}^\mu_- 
+ \left( \partial_\mu \pi_X - g_X X_\mu - {\cal A}_\mu^3 \right)^2
\right] \,.
\label{Lgen2}
\end{align}
Here we have introduced the gauge-invariant combinations
\begin{align}
{\cal A}_\mu^a &= -2 \, {\rm tr} \left[e^{-i \pi_{D_R}} \, 
(i\partial_\mu + g_{D_R} A_{D_R,\mu}) \, e^{i \pi_{D_R}} \, T^a \right]
\simeq  \partial_\mu \pi_{D_R}^a - g_{D_R} A_{D_R,\mu}^a
\end{align}
and 
\begin{align}
 {\cal A}_\mu^\pm &= \frac{1}{\sqrt2} \left( {\cal A}_\mu^1 \pm i {\cal A}_\mu^2 \right) \,.
\end{align}

\subsection{Integrating out the Heavy Spurion Field $\eta_{13}$}

In the standard scenario for the sequence of flavour symmetry breaking discussed in \cite{Feldmann:2009jung}, the next spurion to get a VEV is $\eta_{13}(x)\equiv
\Lambda F_{13} \tilde \eta_{13}(x)$ which
is related to fluctuations around
the CKM angle $\theta_{13} \sim \lambda^3$. We assume that
the spurion potential will generate a mass term for $\eta_{13}$ with a 
generic size of order
\begin{align}
 m_{13}^2 \sim y_b^2 \Lambda^2 \theta_{13}^2 \,.
\end{align}
In the following, we will also
assume that the spurion contribution to $F_X$ in (\ref{FXres}) is
dominating\footnote{Otherwise we would have to integrate out the gauge
boson $X^\mu$ first.}
over the anomaly contribution (\ref{anomcontr}), such that
\begin{align} & 
 \Lambda \gg m_{13} \sim F_X > M_X \,.
\end{align}
We are now going to integrate out the field $\eta_{13}(x)$,
in order to construct the theory below energy scales of order
$F_X$. This in general will induce higher-dimensional operators
for flavour transitions. The leading (tree-level) effects can be 
obtained by solving the equations of motion for $\eta_{13}$,
using the approximate form of the couplings to fermions 
in (\ref{YUapprox},\ref{YDapprox}), leading to
\begin{align}
 {\cal L}_{13} &= \frac12  
    \left(\partial^\mu \eta_{13}\right)^2
- \frac12  \, m_{13}^2  \eta_{13}^2 
- (J_U^{13}+\bar J_U^{13} + J_D^{13} + \bar J_D^{13}) \, \eta_{13}
\,,
\end{align}
with
\begin{align}
 J_U^{13} & \simeq \frac{F_{13}}{2}
\, (\bar U_L',\, \bar D_L' V_{\rm CKM}^\dagger) \,  \frac{\tilde H}{\Lambda}
\left( \begin{array}{ccc}
0 & \frac{\theta_{23} y_b^2 y_c }{y_b^2 y_c^2-y_s^2y_t^2}   &
    \frac{1}{y_t}
  \\[0.3em]
0 & 0 & \frac{\theta_{12} y_s^2 y_t}{y_b^2 y_c^2-y_s^2y_t^2}
 \\[0.3em]
0  & 0  & 0
\end{array} \right)  U_R \,,
\nonumber \\[0.2em]
 J_D^{13} & \simeq  \frac{F_{13}}{2} \, (\bar U_L' V_{\rm CKM},\, \bar D_L') \, 
 \frac{H}{\Lambda} 
\left( \begin{array}{ccc}
0 & -\frac{\theta_{23} y_s y_t^2}{y_b^2y_c^2-y_s^2y_t^2}   &
    -\frac{1}{y_b}
  \\[0.3em]
0 & 0 & - \frac{\theta_{12} y_b y_c^2}{y_b^2y_c^2-y_s^2y_t^2}
 \\[0.3em]
0  & 0  & 0
\end{array} \right) 
 D_R \,,
\label{JUJD}
\end{align}
where the primed fields denote the mass eigenstates of the quarks.
In the limit $m_{13} \to \infty$, 
this leads to an effective 4-fermion interaction
\begin{align}
 \frac{1}{2m_{13}^2} \left( J_U^{13}+\bar J_U^{13} + J_D^{13} +\bar J_D^{13} \right)^2 \,.
\end{align}
Replacing the SM Higgs field $H$ by its VEV in (\ref{JUJD}), 
the effective 4-quark interaction receives an overall suppression
factor $v^2/\Lambda^2$. 
The individual coefficients for specific flavour transitions 
follow from the matrix structure in $J_U^{13}$ and $J_D^{13}$: We observe 
that the $\eta_{13}$ spurion dominantly induces transitions between
left-handed quarks from the first generation and right-handed
quarks from the second or third generation. 
Of course,  more operators -- which may
also include additional gauge fields -- will be generated by
radiative corrections and higher-dimensional operators from 
${\cal L}_{\rm spurion}$ in (\ref{Lspurion}).

\subsection{Integrating out the Heavy $U(1)_X$ Gauge Field}

Below the scale $M_X = g_X F_X$, we may integrate out the heavy
gauge boson of the $U(1)_X$ flavour symmetry and end up with
an effective theory which has an
$$
  [SU(2)_{D_R}] \times U(1)_{u_R} \times U(1)_{D_R^{(2)}}
$$
flavour symmetry. Focusing on the leading terms in (\ref{Lgen2}),
and considering unitary gauge ($\pi_X=0$), we may again solve
the classical e.o.m.\ following from
\begin{align}
{\cal L}_X &\simeq  {\cal L}_{\rm mass}
+ g_X \, X_\mu \, J_X^\mu  \,,
\qquad 
J_X^\mu \equiv \left[ \bar\psi_L \, \gamma^\mu \, Q_X \, \psi_L \right]\,,
\end{align}
in the limit $F_X \to \infty$, where ${\cal L}_{\rm mass}$ is defined
in (\ref{Lgen2}) and $F_X$ in (\ref{FXres}).
Again, this induces effective 4-quark operators
of the form
\begin{align}
 - \frac{1}{2 F_X^2} \left[\bar \psi_L \, \gamma_\mu \, Q_X \, \psi_L \right]
   \left[ \bar\psi_L \, \gamma^\mu \, Q_X \, \psi_L \right] \,,
\end{align}
where $\psi_L$ denotes the set of left-handed fermion fields in
(\ref{psiL}) with the corresponding $U(1)_X$ charges.
The $U(1)_X$ charge operator is diagonal with respect to
the fermion representations. As the up-type quarks Yukawa matrix
is already diagonal, the above operator does not induce FCNCs between
up-type quarks. On the other hand, rotating
the down-type quarks into the mass eigenbasis, one obtains 
\begin{align}
J_X^\mu \Big|_{\rm down} 
& =  \left(\bar d_L',\bar s_L',\bar b_L'\right)
  \gamma^\mu \, X_{D_L}' \left( \begin{array}{c} d_L' \\ s_L' \\ b_L' \end{array}\right)
 \,,
\cr 
X_{D_L}' & = V_{\rm CKM}^\dagger \, {\rm diag}\left[\frac23,-\frac13,-\frac13\right] \, V_{\rm CKM} 
\simeq \left( \begin{array}{ccc}
        \frac{2}{3} & \theta_{12} & \theta_{13} \\
        \theta_{12} & -\frac13 & 0 \\
        \theta_{13} & 0 & -\frac13        
               \end{array}
 \right) + {\cal O}(\theta_{ij}^2)\,,
\label{Xp}
\end{align}
containing FCNCs between $d_L$ and $s_L$ or $b_L$, which are suppressed by
the SM CKM angles.
The phenomenology induced by these sub-leading effects
is qualitatively similar to $Z'$-models with non-universal flavour
couplings \cite{Langacker:2000ju}, where interesting new flavour effects
have been identified in the context of present puzzles in flavour observables
(see e.g.\ \cite{Mohanta:2008ce,Barger:2009eq,Chang:2009wt} for recent applications).
However, compared to the commonly favoured $Z'$ scenarios, our case displays
important modifications:
\begin{itemize}

\item Typical $Z'$-scenarios are motivated by electroweak physics
and consider $Z'$-masses in the TeV range. In this case, precision flavour 
observables in the kaon sector already disfavour non-universal flavour
couplings for the first and second generation.
In our case, the $U(1)_X$ gauge boson is naturally allowed to be much heavier. 
At the same time, the non-universal effects are precisely between the
first and second (or third) generation, and therefore kaon observables 
essentially will provide a lower bound on the scale $F_X$.

\item The $U(1)_X$ gauge boson does not couple to leptons, and thus constraints
from lepton-flavour violating observables do not apply in our case.

\end{itemize}

Taking into account the sub-leading effects proportional to $y_s^2$
in (\ref{Lgen2}), the mixing
between the gauge boson $X_\mu$ with the $SU(2)_{D_R}$ gauge field $A_\mu^3$
induces an additional effective operator, such that finally
\begin{align}
 \label{eq:Leff2} 
{\cal L}_{\rm mass} + g_X X_\mu \, J_X^\mu 
&\to
\frac{y_s^2 \Lambda^2}{4} 
\left[ 
2 {\cal A}_\mu^+ {\cal A}^\mu_- + ({\cal A}_\mu^3)^2 
\right]
- \frac{1}{2 F_X^2} \left[ J_X^\mu \right]^2
-
\frac{y_s^2}{4 \theta_{13}^2 y_b^2} \, J_X^\mu \, {\cal A}_\mu^3 \,.
\end{align}

\subsection{Local $SU(2)_{D_R}$ Flavour Symmetry}

\label{sec:local}

We are now going to repeat our analysis for the effective theory
below the scale $M_X$, where the residual flavour symmetry is
given by
\begin{align}
 G_F' = [SU(2)_{D_R}] \times U(1)_{u_R} \times U(1)_{D_R^{(2)}} \,,
\end{align}
and the $SU(2)_{D_R}$ symmetry is gauged. Since
in this case, the $SU(2)_{D_R}$ is not anomalous, the discussion
is somewhat simpler than in the previous section.
The effective Lagrangian can be obtained as before by combining
kinetic and gauge fixing terms for the $SU(2)_{D_R}$ gauge fields
with the associated Goldstone bosons, the couplings of the gauge fields
to the fermions,
\begin{align}
\label{JD}
 g_{D_R}  A_\mu^a \, (J^\mu_D)^a &= g_{D_R}  A_\mu^a 
\left[ \bar D_R \, \gamma^\mu T^a \, D_R \right] \,,
\end{align}
as well as the kinetic terms and Yukawa couplings of the remaining
spurion fields.

Again, we assume that via an appropriate spurion potential,
the spurion field $y_s(x)=y_s + \eta_s(x)/\sqrt2$ takes its VEV
at a scale of the order of its mass $m_{\eta_s} \sim y_s \Lambda$.
Integrating out the spurion fluctuation $\eta_s(x)$, the Yukawa
coupling to the down-type quarks induces an effective 4-quark operator,
\begin{align} 
\frac1{4m_{\eta_s}^2} \, J_s^2 
\,, \qquad J_s = \frac{1}{\Lambda} \left[
(\bar u_L' \, V_{us} + \bar c_L' \, V_{cs},\, \bar s_L') \, H \, s_R + \rm h.c. \right] \,,
\end{align}
where, again, we have transformed to the quark mass eigenbasis. As
expected, the fluctuations $\eta_s(x)$ around the Yukawa eigenvalue $y_s$ do
not induce flavour transitions, once the SM Higgs VEV is replaced by its VEV.
On the other hand, the charged Goldstone modes in the SM Higgs field, induce
flavour transitions suppressed by the corresponding CKM elements,
in accordance with the principle of MFV.

Starting from (\ref{eq:Leff2}) and (\ref{JD}), we may next integrate out
the $SU(2)_{D_R}$ gauge fields $A_\mu^a$, which acquire a mass
from (\ref{Lgen2}),
\begin{align}
 m_A^2 &= g_{D_R}^2 F_{D_R}^2 \simeq \frac{y_s^2 \Lambda^2}{2} \,.
\end{align}
Taking also into account the leading term from the mixing between $X_\mu$
and $A_\mu^3$, we obtain new effective 4-quark operators
\begin{align}
 \label{Leff3}
- \frac{1}{2 F_{D_R}^2} \left\{ 
   \left[(J_\mu^D)^a\right]^2 
-\frac{ y_s^2}{2 \theta_{13}^2 y_b^2}
   \left[(J_\mu^D)^3 \, J_X^\mu \right] \right\} \,.
\end{align}
Upon Fierz-rearrangement, the term $\left[(J_\mu^D)^a\right]^2$
only involves flavour-diagonal currents $(\bar d_R \gamma_\mu d_R)$ 
and $(\bar s_R \gamma_\mu s_R)$, with different colour structures. 
In the second term, flavour transitions
appear as before, when the current $J_X$ is written in the mass
eigenbasis (\ref{Xp}).

Finally, we may integrate out the field $\eta_{12}$ which we assume
to have a mass of generic order\footnote{
The scalar sector at this stage still contains the fluctuations
around the CP-phase $\delta$ in the CKM matrix which is unrelated
to any of the local or global flavour symmetries, see the discussion
in \cite{Feldmann:2009jung}. In our derivation of (\ref{YUapprox},\ref{YDapprox}),
we have neglected the effects from $\delta$ for simplicity.}
$$
  m_{12}^2 \sim y_s^2 \theta_{12}^2 \Lambda^2 \,.
$$
Notice that the corresponding contributions to the $SU(2)_{D_R}$ gauge
boson masses are sub-leading, and have been neglected in the above analysis
(they lead, however, to small mass splittings between ${\cal A}_\mu^\pm$ 
and ${\cal A}_\mu^3$). In complete analogy to the case of $\eta_{13}$,
see (\ref{JUJD}), we
obtain effective 4-quark operators
\begin{align}
 \frac{1}{2m_{12}^2} \left( J_U^{12}+\bar J_U^{12} + J_D^{12} +\bar J_D^{12} \right)^2 \,,
\end{align}
with
\begin{align}
 J_U^{12} & \simeq \frac{F_{12}}{2}
\, (\bar U_L',\, \bar D_L' V_{\rm CKM}^\dagger) \,  \frac{\tilde H}{\Lambda}
\left( \begin{array}{ccc}
0 & \frac{1}{y_c}   &
    - \frac{\theta_{23} y_b^2 y_t}{y_b^2y_c^2-y_s^2y_t^2}
  \\[0.3em]
0 & 0 & \frac{\theta_{13} y_b^2 y_t}{y_b^2y_c^2-y_s^2y_t^2}
 \\[0.3em]
0  & \frac{\theta_{13} \, y_b^2y_c}{y_b^2 y_c^2-y_s^2 y_t^2}   & 0
\end{array} \right)
  U_R \,,
\nonumber \\[0.2em]
 J_D^{12} & \simeq  \frac{F_{12}}{2} \, (\bar U_L' V_{\rm CKM},\, \bar D_L') \, 
 \frac{H}{\Lambda} 
\left( \begin{array}{ccc}
0 & -\frac{1}{y_s}   &
    \frac{\theta_{23} y_b y_t^2}{y_b^2y_c^2-y_s^2y_t^2}
  \\[0.3em]
0 & 0 & - \frac{\theta_{13} y_b y_t^2}{y_b^2y_c^2-y_s^2y_t^2}
 \\[0.3em]
0  & -\frac{\theta_{13} \, y_s y_t^2}{y_b^2 y_c^2-y_s^2 y_t^2}   & 0
\end{array} \right) 
 D_R \,,
\label{JUJD12}
\end{align}
from (\ref{YUapprox},\ref{YDapprox}).


\section{Global $U(1)_{u_R} \times U(1)_{d_R}$ Flavour Symmetry}

After integrating out the gauge bosons of the local flavour symmetry
and the related spurion fluctuations below the scale $y_s \lambda  \Lambda$, 
the effective theory still possesses
a global $U(1)_{u_R} \times U(1)_{d_R}$ flavour symmetry which acts on
the right-handed quarks of the first generation,
\begin{align}
 u_R & \to e^{i\theta_u} \, u_R \,, \qquad
 d_R \to e^{i\theta_d} \, d_R \,.
\end{align}
We are then left with two complex spurion fields $Y_U^{(1)}$ and
$Y_D^{(1)}$ which break $U(1)_{u_R}\times U(1)_{d_R}$ and give
masses to the lightest quarks ($m_u$, $m_d$).

\subsection{Effective Lagrangian and the Strong CP Problem}
\label{sec:axion}
Let us again parameterize the spurion fields in terms of real
fields representing magnitude and phase,
\begin{align}
  Y_U^{(1)}(x) & = y_u(x) \cdot e^{ - i \pi_u(x)}  \,, \qquad
  Y_D^{(1)}(x)  = y_d(x) \cdot e^{ - i \pi_d(x)}   \,,
\end{align}
where the Goldstone fields in the exponentials transform
as
\begin{align}
 \pi_u &\to \pi_u + \theta_u \,, \qquad 
 \pi_d \to \pi_d + \theta_d \,.
\end{align}
Because of the above shift symmetry of the Goldstone modes,
the (classical) scalar potential only depends on $y_u(x)$ and $y_{d}(x)$,
\begin{align}
 V_0 &= V_0(y_u,y_d) \,.
\end{align}
However,
the non-trivial topological properties of the QCD gauge configurations
imply a more complicated QCD vacuum state beyond perturbation theory. 
The distinct QCD vacua can be labeled by a parameter $\theta$,
and the vacuum-to-vacuum transition amplitude in the $\theta$-vacuum is given by
\begin{align}
 \langle 0|0\rangle_\theta & =
 \sum_{q=-\infty}^\infty \int ({\cal D} A_\mu)_q 
 \int {\cal D} \phi \, \exp[i \, q \, \theta] \, 
  \exp \left[ i \int d^4x \, {\cal L}(A,\phi) \right] \,,
\end{align}
where $\phi$ denotes generic matter fields,
and for a given topological sector $q$, the functional integration 
is restricted to  QCD gauge-potential\footnote{In the following, we concentrate
on the strong-interaction gauge sector, and suppress the weak-interaction
effects in the notation.} configurations $({\cal D} A_\mu)_q$
which satisfy
\begin{equation}
q=\frac{g_s^2}{32 \pi^2} \, \int d^4x \, G_{\mu \nu}^a {\tilde G}^{a \mu \nu} \,.
\end{equation}
The $\theta$-term can thus be considered as being part of
an effective term in the Lagrangian 
\begin{equation}
 {\cal L} \to {\cal L}_{\rm eff}={\cal L}+ \, \theta \, \frac{g_s^2}{32 \pi^2} \, G_{\mu \nu}^a {\tilde G}^{a \mu \nu} \,.
\end{equation}

Compared  to the SM case, 
the additional chiral $U(1)_{u_R} \times U(1)_{d_R}$ flavour symmetry
now allows for a solution of the strong CP problem
via a Peccei/Quinn \cite{Peccei:1977ur,Weinberg:1977ma,Wilczek:1977pj} mechanism.\footnote{The 
connection between flavour and PQ symmetries has been discussed
before, see e.g.\ \cite{Wilczek:1982rv,Berezhiani:1989fp,Cheung:2010hk}.}
Here we recall that the effective action (in the QCD gauge sector)
changes under chiral rotations as 
\begin{eqnarray}
 \Gamma &\to & \Gamma + q \, (\theta_u + \theta_d) \,,
\end{eqnarray}
which is equivalent to a change in the QCD $\theta$-parameter,
$$
 \theta \rightarrow \theta - \theta_{u} - \theta_{d} \,.
$$
For $\langle y_{q} \rangle >0$
(by assumption), the fermion mass term gets its canonical form, 
after a chiral transformation of $u_R$ and $d_R$ with
the corresponding phases set by $\langle \pi_u(x)\rangle$
and $\langle \pi_d(x) \rangle$, respectively.
To avoid the strong CP problem, one thus has to require that
\begin{eqnarray}
\langle \theta_{\rm eff} \rangle \equiv \theta - \langle \pi_u(x) + \pi_d(x) \rangle \stackrel{!}{=} 0 \,. 
\label{thetaeff}
\end{eqnarray}
This can be achieved by examining the effective potential in the
presence of the QCD $\theta$-vacuum which --- for small VEVs 
$\langle y_{u,d} \rangle  \ll 1$ --- takes the 
general form (see appendix and \cite{Peccei:1977ur})
\begin{align}
 V_\theta = V_0 
 - K \, v^6 \, y_c y_t y_{s} y_{b} \ y_u(x) \, 
   y_{d}(x) \, \cos\left[\pi_u(x) + \pi_d(x) - \theta\right]  + \ldots
\label{V2}
\end{align}
with a positive constant $K>0$.
The potential (\ref{V2}) thus 
breaks the original shift symmetry for the Goldstone fields.
Its minimum is given by 
$\langle \pi_u + \pi_d \rangle = \theta$,
and therefore $\langle \theta_{\rm eff} \rangle \equiv 0$, as required.
Notice that the potential only depends on the combination 
\begin{align} 
a(x) &\equiv f_a \left( \pi_u(x)+\pi_d(x) \right) \,,  
\end{align}
which defines the PQ axion field,
with $f_a$ being a dimensional normalization constant. 
The corresponding PQ symmetry is defined such that
the axion transforms as 
\begin{align}
 a(x) & \to a(x) + f_a \, \theta_{\rm PQ} \,. 
\end{align}
In order to determine the normalization $f_a$ and to find the
orthogonal linear combination of $\pi_u(x)$ and $\pi_d(x)$, which we
denote as $b(x)$, we 
consider the flavour invariant kinetic term and require
\begin{align}
\Lambda^2 \,  \partial_\mu Y_U^{(1)} \partial^\mu Y_U^{(1)\dagger }
+
\Lambda^2 \, \partial_\mu Y_D^{(1)} \partial^\mu Y_D^{(1)\dagger } \, \Bigg|_{y_{u,d} \to \langle y_{u,d} \rangle} 
& \stackrel{!}{=} \frac12 \left(\partial_\mu a(x)\right)^2
 + \frac12 \left(\partial_\mu b(x)\right)^2 \,.
\end{align}
This yields
\begin{align}
 f_a &= \sqrt2 \,  \Lambda \, \left\langle \frac{y_d y_u}{\sqrt{y_d^2 + y_u^2}} \right\rangle  \,,
\qquad 
 b(x) = f_a \left( \left\langle \frac{y_u}{y_d}\right\rangle  \, \pi_u(x) 
           - \left\langle \frac{y_d}{y_u} \right\rangle \, \pi_d(x) \right) \,.
\end{align}
In terms of $a(x)$ and $b(x)$, the up- and down-quark Yukawa couplings read
\begin{align}
Y_U^{(1)}(x) &= \exp\left[ - i \,  \left\langle \frac{y_d^2}{y_u^2+y_d^2} \right\rangle \, \frac{a(x)}{f_a}
  \right] \exp\left[ - i \,  \left\langle \frac{y_u y_d}{y_u^2+y_d^2} \right\rangle \, \frac{b(x)}{f_a}
  \right]  y_u(x) \,,
\cr 
Y_D^{(1)}(x) &= \exp\left[ - i \,  \left\langle \frac{y_u^2}{y_u^2+y_d^2} \right\rangle \, \frac{a(x)}{f_a}
  \right] \exp\left[ + i \,  \left\langle \frac{y_u y_d}{y_u^2+y_d^2} \right\rangle \, \frac{b(x)}{f_a}
  \right]  y_d(x) \,.
\end{align}
We also define the corresponding linear combinations of
$U(1)$ charges,
\begin{align}
 & \theta_{\rm PQ} = \theta_u+\theta_d \,, \qquad
 \theta_{\rm diff} = \left\langle\frac{y_u}{y_d}\right\rangle \, \theta_u 
   - \left\langle \frac{y_d}{y_u} \right\rangle \, \theta_d
\,, 
\end{align}
such that the orthogonal combination of Goldstone bosons
transforms as $$b(x) \to b(x) + f_a \theta_{\rm diff} \,.$$
Note that $b(x)$ remains massless, apart from anomalous contributions from 
the electroweak vacuum. We may or may not remove $b(x)$ by gauging the 
remaining $U(1)_{\rm diff}$ symmetry and subsequently integrating out the
corresponding massive gauge boson. 

The axion field remains in the physical spectrum of the theory. However, compared 
to the original Peccei-Quinn axion, its couplings are now determined by the scale 
$\Lambda$ of the Yukawa fields and not by the electroweak VEV. In particular, the scale 
$\Lambda$ has to be chosen well above the electroweak scale, in which case the 
axion couplings become very small, since they scale as $1/f_a$. Thus the phenomenology 
of this model will be similar to the phenomenology of invisible axion models 
\cite{Kim:1979if,Shifman:1979if}. 


\section{Conclusions}
In this paper, we have discussed the scenario of a spontaneously broken flavour 
symmetry, realized in a set-up with Standard Model (SM) fermion representations, 
and standard Yukawa matrices which are promoted to dynamical scalar fields
that are subject to an appropriately chosen scalar potential. 
In order to avoid
massless Goldstone bosons in the physical spectrum,
we consider \emph{local} flavour symmetries, where the corresponding
gauge bosons become massive by the usual Higgs mechanism.
Our scenario necessarily has to be understood in the context of an
effective theory approximation for a more fundamental theory.
First, before the scalar fields are integrated out,
the Yukawa interactions are now described by dimension-5 operators.
Second, the flavour symmetries of the SM are anomalous, destroying
the local gauge symmetry of the classical Lagrangian. Also,
the mixed anomalies with the electroweak symmetry could be problematic.
Following \cite{Preskill:1990fr}, 
we have pointed out that the local symmetry can be formally restored by adding appropriate
higher-dimensional operators involving the Goldstone fields, which can
be viewed as the remnant of integrating out very heavy fermion representations
which cancel the anomalies in the fundamental theory. 
The effect of the anomalies can then be absorbed into the masses of the heavy gauge 
bosons of the broken flavour symmetry, and the mixed anomalies can be removed 
by choosing the counter terms appropriately.

The masses of the new heavy gauge bosons as well as of the 
new physical Higgs modes are hierarchically ordered, according
to the sequence of hierarchies in the quark masses and mixing angles worked 
out in \cite{Feldmann:2009jung}.
The masses of the lightest of these new states therefore have to be 
sufficiently large, such that the induced flavour-changing transitions 
are in line with the experimental constraints from precision measurements
in the kaon and $B$-meson sector.
We have particularly concentrated on 4-quark
operators which appear after integrating out the heavy gauge bosons and 
scalar fields at tree-level. We found -- not surprisingly 
-- that our set-up shares the basic features of minimal flavour violation
where all non-trivial flavour structures are induced by the SM Yukawa
couplings and CKM elements. We have discussed examples where the 
flavour effects induced by integrating out the new heavy gauge bosons 
share some features of $Z'$-models with
non-universal flavour couplings. On the other hand, the scalar fluctuations 
around the VEV of the CKM angles directly lead to flavour-changing neutral currents in
the effective low-energy theory, which may be checked experimentally.
We shall work out 
the phenomenological implications of this scenario 
for flavour observables accessible by future experiments 
at the LHC or at Super-B factories in a separate publication. 

Finally, we have entertained the possibility to leave the chiral $U(1)$ factors
in the SM flavour symmetry group as a global Peccei-Quinn symmetry, which
allows to avoid the strong CP-problem when the Goldstone modes dynamically
lead to a vanishing effective  $\theta$-parameter in QCD. The couplings
of the physical axion field in such a scenario is strongly suppressed
by the UV scale of the effective theory which is also responsible for 
the small Yukawa interactions of the first quark family.

In summary, while (admittedly) still being a rather speculative new physics scenario,
our set-up provides an interesting alternative
to generate the observed hierarchies in the quark masses and mixings, with well-defined physical
consequences. We should also remark that -- in contrast to many other generic new physics models --
our approach is protected by MFV against unacceptably large corrections to flavour observables.

\section{Acknowledgements}
TM and TF like to thank Aneesh Manohar for useful discussions. 
MA has been supported by the German Research Foundation
(DFG, GRK~1054).
TM has been supported in part by the German Research Foundation 
(DFG, Contract No.~MA1187/10-1), 
by the German Ministry of Research (BMBF, Contract No.~05HT6PSA), and by the 
European Union, MRTN-CT-2006-035482 (FLAVIAnet).

\begin{appendix}

\section{Anomalous $U(1)_X$ Theory}

\label{app:U1X}

The general formalism for a consistent formulation of
gauge theories based on an anomalous local symmetry within an
effective-theory framework has been given in \cite{Preskill:1990fr}.
In our case, we have to deal with an anomalous $U(1)_X$ symmetry,
which acts on the chiral fermion representations in the SM according
to (\ref{eq:Xdef}).
The classical part of the Lagrangian involving the $U(1)_X$ gauge field
reads
\begin{eqnarray}
 {\cal L}_X &=& - \frac14 \, X_{\mu\nu}(x) X^{\mu\nu}(x) \,, \\
 {\cal L}_\psi &=& \bar\psi_L(x) \, i \Slash D \, \psi_L(x)  \quad \mbox{and} \quad
 {\cal L}_{\rm spurion}
\end{eqnarray}
where $D_\mu= \partial_\mu - i g_X q_X X^\mu + \ldots $, and we have
only shown the coupling to the $U(1)_X$ gauge boson 
in the covariant derivative including the $U(1)_X$ charges 
for the different fermion species in (\ref{psiL}).
While the classical theory is invariant under the gauge transformation 
\begin{eqnarray}
 X_\mu(x) &\to& X_\mu(x) + \frac{1}{g_X} \, \partial_\mu \omega_X(x) \,,
\end{eqnarray}
which can be compensated by a local phase transformation of
the fermions,
\begin{eqnarray}
 \psi_L(x) &\to& e^{i \omega_X(x) \, q_X} \, \psi_L(x) \,,
\end{eqnarray}
in the classical Lagrangian, the quantum effective action changes 
since the representation $\psi_L$ of $U(1)_X$ is anomalous,
\begin{align}
{\rm tr} \left[ Q_X^3 \right] = \frac{3}{4} \neq 0 \,.
\end{align}
We note that, due to the fact that the charge $Q_X$ is a linear
combination of $SU(3)^3$ generators, its trace vanishes ${\rm tr}[Q_X]=0$ 
individually for every SM gauge multiplet,
and therefore we will not encounter mixed anomalies with $SU(N)$ generators
(where ${\rm tr}[\{ T^a,T^b \}  Q_X ] \propto \delta^{ab} {\rm tr}[Q_X]=0$) 
or gravitons. Also ${\rm tr}[Y^2  Q_X]=0$. 
There are, however, mixed anomalies with the SM hypercharge from triangle
diagrams involving two $U(1)_X$ gauge fields and one hypercharge gauge field,
\begin{align}
 {\rm tr}[Q_X^2 \, Y] = -1 \neq 0 \,,
\end{align}
(where we normalized the hypercharge such that $Y(Q_L)=1/3$, $Y(U_R)=4/3$, $Y(D_R)=-2/3$).
This implies that the total change of the effective action under 
either $U(1)_X$ or $U(1)_Y$ local transformations is given by \cite{Preskill:1990fr}
\begin{eqnarray}
 \delta_{\omega_X} \Gamma &=& {\rm tr}[Q_X^3] \, \frac{g_X^2}{48\pi^2}
 \, \int d^4x \, \omega_X \, X_{\mu\nu} \, \tilde X^{\mu\nu} 
 \cr && {} + c_1 \, {\rm tr}[Q_X^2 \, Y] \, \frac{g_Y g_X}{48\pi^2} 
\int d^4x \, \omega_X \, X_{\mu\nu} \, \tilde Y^{\mu\nu} 
\,, \\
\delta_{\omega_Y}\Gamma &=& (1-c_1) \, {\rm tr}[Q_X^2 \, Y] \, \frac{g_X^2}{48\pi^2} 
\int d^4x \, \omega_Y \, X_{\mu\nu} \, \tilde X^{\mu\nu} \,,
\end{eqnarray}
where 
\begin{align}
 \tilde X^{\mu\nu} = \frac12 \, \epsilon^{\mu\nu\rho\sigma} \, X_{\rho\sigma}
\end{align}
is the field-strength dual.
The coefficient $c_1$ arise from the freedom to add
an appropriate local counter-term \cite{Preskill:1990fr},
\begin{equation}
 \Gamma_{\rm c.t.} = c_1 \, {\rm tr}[Q_X^2 \, Y] \, \frac{g_X^2 g_Y}{24\pi^2} 
 \, \int d^4x \, \epsilon_{\mu\nu\lambda\sigma} \, Y^\mu
 X^{\nu} \partial^\lambda X^{\sigma} \,,
\end{equation}
which changes under gauge transformations as
\begin{eqnarray}
 \delta_{\omega_Y}\Gamma_{\rm c.t.} &=& 
 - c_1 \, {\rm tr}[Q_X^2 \, Y] \, \frac{g_X^2}{48\pi^2} 
\int d^4x \, \omega_Y(x) \, X_{\mu\nu}  \tilde X^{\mu\nu} 
\,, \\
\delta_{\omega_X}\Gamma_{\rm c.t.} &=& c_1\, {\rm tr}[Q_X^2 \, Y] \, \frac{g_X \, g_Y}{48\pi^2} 
\int d^4x \, \omega_X(x) \, X_{\mu\nu}  \tilde Y^{\mu\nu} \,.
\end{eqnarray}
Choosing the renormalization condition $c_1=1$, the $U(1)_Y$ symmetry remains 
manifestly non-anomalous, $\delta_{\omega_Y}\Gamma =0$, whereas
\begin{eqnarray}
 \delta_{\omega_X} \Gamma &=& 
\frac{1}{48\pi^2} \int d^4x \, \omega_X \left\{
{\rm tr}[Q_X^3] \, g_X^2  X_{\mu\nu}  \tilde X^{\mu\nu}
+  {\rm tr}[Q_X^2 \, Y] \, g_X g_Y   X_{\mu\nu}  \tilde Y^{\mu\nu} 
\right\}
\,.
\end{eqnarray}
Still, local gauge invariance can be formally restored by 
exploiting the behaviour of the Goldstone field $\pi_X(x)$
under gauge transformations,
\begin{equation}
 \pi_X(x) \to \pi_X(x) + \omega_X(x) \,.
\label{shift}
\end{equation}
Adding a term
\begin{align}
\Delta {\cal L}_{\rm \pi} & = 
 - \ \frac{\pi_X(x)}{48\pi^2}
\left\{
{\rm tr}[Q_X^3] \, g_X^2  X_{\mu\nu}(x)  \tilde X^{\mu\nu}(x)
+  {\rm tr}[Q_X^2 \, Y] \, g_X g_Y   X_{\mu\nu}(x)  \tilde Y^{\mu\nu}(x) 
\right\}
 \,,
\label{eq:Lpi}
\end{align}
then compensates for the change in $\Gamma$ from the fermion measure.
On the quantum level, loop corrections involving the anomalous coupling of 
the Goldstone mode $\pi_X$ to the gauge fields in (\ref{eq:Lpi})
will also lead to a mass term for the $U(1)_X$ gauge boson \cite{Preskill:1990fr}
(in addition to the masses generated by the spurion VEVs in (\ref{Lgen1},\ref{Lgen2})).
These diagrams are quadratically divergent and contribute as
\begin{eqnarray}
 M_X \equiv g_X F_X \ \ni \ \frac{g_X^3 \, {\rm Tr}[Q_X^3]}{64\pi^3} \, \Lambda \,,
\mbox{\small \quad respectively \quad } 
 \frac{g_Y g_X^2 \, {\rm Tr}[Y Q_X^2]}{64\pi^3}  \, \Lambda 
\label{anomcontr}
\end{eqnarray}
to the $U(1)_X$ gauge boson mass, with 
the corresponding quadratic term in the effective Lagrangian
\begin{eqnarray}
 {\cal L}_{\rm \pi} &=& 
   \frac{F_X^2}{2} \, \left(\partial_\mu \pi_X(x) - g_X X_\mu(x)\right)^2 \,.
\label{eq:Mpi}
\end{eqnarray}
Here $F_X$ is a dimensional constant such that $(F_X \pi_X)$
has canonical dimensions and a correctly normalized kinetic term.
The leading contribution to $F_X$ from the spurion-induced spontaneous
symmetry breaking can be read off (\ref{Lgen2}),
\begin{align}
 F_X^2 &= 2 y_b^2 \theta_{13}^2  \Lambda^2 + \ldots
\label{FXres}
\end{align}
Covariant gauges can be introduced via the gauge fixing term
\begin{equation}
 {\cal L}_{\rm g.f.} = - \frac{1}{2\alpha_X} \left(\partial_\mu X^\mu(x) - \alpha_X \, g_X \, F_X^2 \, \pi_X(x) \right)^2 \,,
\end{equation}
which in combination with the kinetic term in ${\cal L}_\pi$, gives the
quadratic terms
\begin{eqnarray}
 && \frac{M_X^2}{2} \, (X_\mu)^2 - \frac{1}{2\alpha_X} \, (\partial_\mu X^\mu)^2  
 - \alpha_X \, \frac{M_X^2}{2} \, (F_X \pi_X)^2 
+ \frac{1}{2} \, (F_X \partial_\mu \pi_X)^2 \,.
\end{eqnarray}
The mixing term between $X^\mu$ and $\pi_X$ vanishes (up to a total derivative).
For the 't~Hooft-Landau gauge, $\alpha_X=0$, and the $\pi_X$ field is massless.
In unitary gauge, $\alpha_X \to \infty$, the Goldstone field decouples.

\section{Unitary gauge for the 2--Family Case}

\label{app:unitary2}

In the 2-family case, Eqs~(\ref{cond1}--\ref{cond3}) can be easily solved for the 
various $\xi_X$-fields,
\begin{align}
\xi_{U_L}^1(x) & = 0 \,, \qquad  
\xi_{U_L}^2(x)  = - F_{12}^2 \, \frac{y_u^2+y_c^2}{(y_u^2-y_c^2)^2} \, \eta_{12}(x)
\simeq - \frac{F_{12}^2}{y_c^2} \, \eta_{12}(x) \,, 
\\ 
\xi_{D_L}^1(x) & = 0 \,, \qquad  
\xi_{D_L}^2(x)  = - F_{12}^2 \, \frac{y_d^2+y_s^2}{(y_d^2-y_s^2)^2} \, \eta_{12}(x)
\simeq - \frac{F_{12}^2}{y_s^2} \, \eta_{12}(x) \,, 
\\ 
\xi_{U_R}^1(x) & =  -F_{12}^2 \, \frac{2 y_u y_c \sin\pi_u}{(y_u^2-y_c^2)^2} \,
\eta_{12}(x) \simeq 0
\,, \qquad 
\xi_{U_R}^2(x)  = 
 - F_{12}^2 \, \frac{2 y_u y_c \cos\pi_u}{(y_u^2-y_c^2)^2} \, \eta_{12}(x) 
\simeq 0
\,, \\
\xi_{d_R}^1(x) & \to  + F_{12}^2 \, \frac{2 y_d y_s \sin\pi_d}{(y_d^2-y_s^2)^2}
\, \eta_{12}(x)
\simeq 0
\,, \qquad 
\xi_{d_R}^2(x)  \to  + F_{12}^2 \, \frac{2 y_d y_s \cos\pi_d}{(y_d^2-y_s^2)^2}
\, \eta_{12}(x)
\simeq 0 \,,
\end{align}
leaving one dynamical spurion field $\eta_{12}(x)$ which describes fluctuations
around the Cabibbo angle.
The factor
\begin{align}
F_{12}^2 & = \frac{2 (y_u^2-y_c^2)^2 (y_d^2-y_s^2)^2}{(y_u^2-y_c^2)^2
  (y_d^2+y_s^2) + (y_d^2-y_s^2)^2 (y_u^2+y_c^2)}
\end{align}
has been introduced to normalize the kinetic term for $\eta_{12}(x)$
in the form
$$
 \frac12 \, F_{12}^2 \, \Lambda^2\, \left(\partial^\mu \eta_{12}(x)\right)^2
$$
when
inserted into (\ref{kin}). It is important to notice that the
fluctuation $\eta_{12}(x)$ appears symmetrically in the up- and in the
down-quark sector (scaled by ratios of quark Yukawa couplings), despite the
fact that our original ansatz (\ref{ansatz}) assigned the CKM matrix solely
to the down-quark Yukawa matrix. In contrast, the naive replacement 
$\theta_{12} \to \theta_{12} + \frac{F_{12}^2 \,\eta_{12}(x)}{\sqrt2}$
in (\ref{Yvev},\ref{Vckm}) would have lead to an incorrect result, where $\eta_{12}(x)$ 
would only couple to down-type quarks.\footnote{Still, in the limit
$y_{u,d} \ll y_s \ll y_c$, the coupling of $\eta_{12}(x)$ 
to $d_L$ and $s_R$ will dominate.}
 Linearizing in the fluctuations
$\eta_{12}(x)$, the Yukawa couplings in unitary gauge
for the 2-family case read
\begin{align}
 Y_U^{\rm u.g.}(x) &= {\rm diag}\left[y_u e^{-i\pi_u} , y_c \right] + 
 \left( \begin{array}{cc}
0 & - \frac{y_c}{2 (y_c^2-y_u^2)} \\
- \frac{y_u e^{-i\pi_u}}{2 (y_c^2 - y_u^2)} & 0
\end{array} \right) F_{12}^2 \, \eta_{12}(x) + {\cal O}(\eta_{12}^2) \,,
\cr 
 Y_D^{\rm u.g.}(x)& = V_{\rm CKM} \cdot \left\{
{\rm diag}\left[y_d e^{-i\pi_d} , y_s \right] + 
 \left( \begin{array}{cc}
0 & \frac{y_s}{2 (y_s^2-y_d^2)} \\
\frac{y_d e^{-i\pi_d}}{2 (y_s^2 - y_d^2)} & 0
\end{array} \right) F_{12}^2 \, \eta_{12}(x) + {\cal O}(\eta_{12}^2)
\right\} \,.
\cr 
\label{fix2}
\end{align}
The coupling matrices of the $\eta_{12}(x)$ field can be expressed 
entirely in terms of the VEVs of the Yukawa matrices,
\begin{align}
& \delta Y_U \equiv F_{12}^2 \, \left( \begin{array}{cc}
0 & - \frac{y_c}{2 (y_c^2-y_u^2)} \\
- \frac{y_u e^{-i\pi_u}}{2 (y_c^2 - y_u^2)} & 0
\end{array} \right)
\cr 
&=\frac{F_{12}^2}{y_c^2-y_u^2}
\left\{ 
- \frac{
 y_c^2 y_s^2 \tan\theta_{12} 
+y_c^2 y_d^2 \cot\theta_{12}
-y_u^2 y_s^2 \cot\theta_{12}
-y_u^2 y_d^2 \tan\theta_{12}}{2 (y_c^2-y_u^2)(y_d^2-y_s^2)} \, \langle Y_U \rangle
\right.
\cr 
& \left. \phantom{\frac{1}{y_c^2-y_u^2}}  \qquad 
+
\frac{\cot2\theta_{12}}{y_c^2-y_u^2} \, \langle Y_U Y_U^\dagger Y_U \rangle
-
\frac{\csc2\theta_{12}}{y_s^2-y_d^2} \, \langle Y_D Y_D^\dagger Y_U \rangle
\right\}
\cr 
& \simeq
\frac{y_s^2 \tan\theta_{12}}{y_c^2} \, \langle Y_U \rangle 
+
\frac{2 y_s^2 \cot2\theta_{12}}{y_c^4} \, \langle Y_U Y_U^\dagger Y_U\rangle 
- \frac{2\csc2\theta_{12}}{y_c^2} \, 
 \langle Y_D Y_D^\dagger Y_U \rangle
\end{align}
with an analogous relation for $\delta Y_D$.
Here the first identity holds in the basis where $\langle Y_U \rangle$
is diagonal and $\langle Y_D \rangle = V_{\rm CKM} D_D$, while the
second and third line are basis independent,
due to the transformation properties of $\delta Y_U \sim (3,\bar 3,1)$
and $\delta Y_D \sim (3,1,\bar 3)$ under the flavour symmetry group.
From the MFV perspective, we thus expect the coefficients in front of
the 3 individual flavour structures to be of ${\cal O}(1)$ or smaller.
Indeed, taking into account that $\eta_{12}(x) \sim \theta_{12}$, we obtain
\begin{align}
& \frac{y_s^2 \tan\theta_{12}}{y_c^2} \, \eta_{12} \sim  y_s^2 \theta_{12}^2 \ll 1 \,,
\quad
\frac{2 y_s^2 \cot2\theta_{12}}{y_c^4} \, \eta_{12} \sim y_s^2 \ll 1 \,,
\quad
 \frac{2\csc2\theta_{12}}{y_c^2} \, \eta_{12} \sim 1 \,,
\end{align}
for $y_c \sim {\cal O}(1)$, $y_c \gg y_s \gg y_{u,d}$.
It is interesting to note that
\begin{align}
 {\rm tr}\left[\langle Y_U^\dagger\rangle \, \delta Y_U \right]
 = {\rm tr}\left[\langle Y_U^\dagger Y_U Y_U^\dagger \rangle \, \delta Y_U \right] = 0 \,,
\end{align}
which shows that our construction for $\eta_{12}(x)$ 
indeed involves a variation that is orthogonal to the VEV of $Y_U$
(an analogous statement holds for $\delta Y_D$ and $\langle Y_D \rangle$).

\section{Peccei--Quinn Mechanism for $U(1)_{u_R} \times U(1)_{d_R}$}

To show that (\ref{thetaeff}) is satisfied,
we follow the original paper by Peccei/Quinn \cite{Peccei:1977ur}
and consider the generating functional (in Euclidean space),
\begin{eqnarray}
 Z_{\theta}(J_{U,D},J_{U,D}^\dagger)&=&
\sum_q e^{i \theta q} \, 
\int ({\cal D} A_{\mu})_q 
 \int {\cal D} Y_{U,D} \, {\cal D} Y_{U,D}^{\dagger} 
\int {\cal D} \Psi  \,{\cal D} \bar \Psi \,
\nonumber \\[0.2em] &\times& 
 \exp\left\{ \int d^4x  \left( {\cal L}(x)
+{\rm tr} [J_U(x) \, Y_U(x)]
+{\rm tr} [J_D(x) \, Y_D(x)]  + \mbox{h.c.} \right) \right\} \,,
\cr &&
 \label{Z} 
\end{eqnarray}
where ${\cal D}\Psi$ denotes all left- and right-handed quarks,
and ${\cal L}(x)$ contains the SM Yukawa terms with the Higgs field
set to its VEV,
\begin{align}
- {\cal L}_{\rm yuk} &= v \, \bar U_L(x) \, Y_U(x) \, U_R(x) 
                      + v \bar D_L(x) \, Y_D(x)\, D_R(x) + \mbox{h.c.} 
\end{align}
The effective potential in the non-trivial QCD vacuum can be 
obtained from an expansion in small Yukawa couplings, with the leading
term coming from the $q=\pm 1$ sectors, leading to
\begin{align}
 V_\theta = V_0 - K \, v^6 \, {\rm Re}\left[ \det Y_U \, \det Y_D \, e^{i \theta} \right]
\end{align}
with a positive constant
\begin{align}
 K = \frac{ \int ({\cal D} A_{\mu})_1 \!
\int {\cal D} \Psi  \,{\cal D} \bar \Psi 
  \,
 e^{\int d^4x \left( -\frac14 GG + \bar\Psi \, i\Slash D \, \Psi \right)}
\, \prod_{i=u,c,t,d,s,b} \, \int d^4x_i\, \bar \Psi_L^i(x_i) \Psi_R^i(x_i) }
{\int ({\cal D} A_{\mu})_1 \!
\int {\cal D} \Psi  \,{\cal D} \bar \Psi   \,
 e^{\int d^4x \left( -\frac14 GG + \bar\Psi \, i\Slash D \, \Psi \right)}}
\end{align}
and $K \sim \Lambda_{\rm QCD}^2$.
After spontaneous symmetry breaking, $G_F \to U(1)_{u_R} \times U(1)_{d_R}$,
the Yukawa matrices are represented as
\begin{align}
 Y_U(x) = \left( \begin{array}{ccc} y_u(x) \, e^{-i\pi_u(x)} & 0 & 0
                  \\ 0 & y_c & 0 \\ 0 & 0 & y_t
                 \end{array} \right)
\,, \qquad 
 Y_D(x) = V_{\rm CKM}
   \left( \begin{array}{ccc} y_{d}(x) \, e^{-i\pi_d(x)} & 0 & 0
                  \\ 0 & y_s & 0\\ 0 & 0 & y_{b}
                 \end{array} \right) \,,
\end{align}
with
\begin{align}
 \det Y_U = y_u y_c y_t e^{-i \pi_u} \,, \qquad 
 \det Y_D = y_{d} y_{s} y_{b} e^{-i \pi_d} \,,
\end{align}
from which (\ref{V2}) follows.

\end{appendix}

\end{document}